\documentclass{Interspeech}

\interspeechcameraready

\title{An Investigation on Speaker Augmentation for End-to-End Speaker Extraction}

\author[affiliation={1}]{Zhenghai}{You}
\author[affiliation={1}]{Zhenyu}{Zhou}
\author[affiliation={1}]{Lantian}{Li}
\author[affiliation={2}]{Dong}{Wang}

\affiliation{School of Artificial Intelligence}{Beijing University of Posts and Telecommunications}{China}
\affiliation{Center for Speech and Language Technologies, BNRist}{Tsinghua University}{China}

\email{lilt@bupt.edu.cn, wangdong99@mails.tsinghua.edu.cn}


\usepackage{comment}

\usepackage{amsmath}
\usepackage{amssymb}

\usepackage{graphicx}
\usepackage{subcaption}
\usepackage{float}

\usepackage{booktabs}
\usepackage{multirow}
\usepackage{makecell}

\begin{document}
\maketitle

\begin{abstract}
Target confusion, defined as occasional switching to non-target speakers, poses a key challenge for end-to-end speaker extraction (E2E-SE) systems. We argue that this problem is largely caused by the lack of generalizability and discrimination of the speaker embeddings, and introduce a simple yet effective speaker augmentation strategy to tackle the problem. Specifically, we propose a time-domain resampling and rescaling pipeline that alters speaker traits while preserving other speech properties. This generates a variety of pseudo-speakers to help establish a generalizable speaker embedding space, while the speaker-trait-specific augmentation creates hard samples that force the model to focus on genuine speaker characteristics. Experiments on WSJ0-2Mix and LibriMix show that our method mitigates the target confusion and improves extraction performance. Moreover, it can be combined with metric learning, another effective approach to address target confusion, leading to further gains. 
\end{abstract}

\section{Introduction}

Extracting a specific speaker's voice from multi-talker speech signals is a fundamental challenge in speech signal processing, commonly referred to as \textbf{S}peaker \textbf{E}xtraction (SE)~\cite{xu2020spex,elminshawi2022new,zmolikova2023neural}. 
Unlike Speaker Separation (SS)~\cite{choi2005blind,naik2014blind,pal2013blind}, 
which aims to separate all speakers in a mixed speech, 
SE refers to an enrollment utterance of the target speaker and selectively extracts the speech of the target speaker.

In recent years, the \textit{end-to-end} speaker extraction (E2E-SE) approach has gained popularity. In this approach, a speaker encoder and a speech encoder are integrated with a speaker extractor, forming a unified end-to-end extraction framework~\cite{xu2020spex,delcroix2020improving,liu2023x, chen2023mc}. 
Typically, the speech encoder produces latent representations of the mixed speech, 
while the speaker encoder generates a target speaker embedding of the reference speech. 
The speaker extractor then extracts the target speaker's speech from the mixed speech representations by referring to the target speaker embedding.

Despite significant advancements, 
the \textit{target confusion} problem remains a challenging issue~\cite{liu2023x,zhao2022target,pan2022hybrid}. 
This problem occurs when the system extracts speech components of the interfering speaker as the target speech, 
resulting in a low-quality listening experience. 
This problem is particularly severe when the target and interfering speakers share similar vocal characteristics, 
making them difficult to distinguish. It also occurs when the mixed speech is severely corrupted by noise, in which case the speaker traits of the target and interfering speech are equally difficult to identify, leading to potential target confusion.


Recently, several approaches have been proposed to address the target confusion problem. 
From the data perspective, Li \textit{et al.}~\cite{li2024effectiveness} introduced noise and reverberation into enrollment utterances during model training, to improve the robustness of the speaker encoder in complex conditions. 
From the model perspective, architectures such as SpEx++~\cite{ge2021multi} and DPRNN-IRA~\cite{2020Robust} 
incorporate an iterative refinement mechanism that incrementally enhances the embeddings of the target speaker. 
From the objective perspective, Zhao \textit{et al.}~\cite{zhao2022target} proposed a metric learning approach, 
which introduces an additional triplet loss to contrast the distance between the enrollment-target pairs and the enrollment-interfering pairs.

Speaker augmentation has been shown to be very effective in speaker recognition tasks~\cite{yamamoto2019speaker,chen2022build,zhou2024comprehensive}.
Briefly, this approach produces some `pseudo speakers' by perturbing the vocal fold or vocal tract properties of some existing real speech utterances, or utilizing multi-speaker text-to-speech or voice conversion techniques. 
This paper explores the possibility of addressing the target confusion problem by speaker augmentation. 
We choose a simple time-domain resampling and rescaling pipeline. 
By this approach, we can modify the fundamental frequency (F0) and formants of the original speech, and this modified speech can be regarded as from `new' speakers or pseudo speakers. Adding the generated speech into the training dataset is supposed to improve the construction of the speaker space, leading to increased generalizability of the speaker embeddings. 
Most importantly, the augmentation changes the speaker traits only, while preserving all the non-speaker properties, e.g., textual content, tempo, and prosody of the original speech.   
We argue that when the augmented version and the original version of the same speech are mixed, it forms a ``hard'' sample for the model training, as it is more difficult to process than usual mixture samples where the target and interfering speech usually possess different content, tempo, and prosody. Our hypothesis is that involving such hard samples in model training will compel the system to capture the genuine speaker characteristics, by which the target confusion problem can be mitigated.

We conduct experiments on two benchmark datasets, WSJ0-2Mix~\cite{hershey2016deep} and LibriMix~\cite{cosentino2020librimix}, 
demonstrating that our method consistently improves performance with two state-of-the-art E2E-SE architectures, 
including DPRNN~\cite{2020Dual} and SpEx+~\cite{ge2020spex+}. 
The ablation study shows that the performance improvement can be largely attributed to the hard samples produced by speaker augmentation. 
Additionally, we show that our method can be combined with other advanced techniques, such as metric learning, resulting in further performance improvement.

The rest of this paper is structured as follows. 
Section~\ref{sec:rela} reviews related work and Section~\ref{sec:meth} details the proposed speaker augmentation approach. 
Section~\ref{sec:exp} presents the experimental results, and Section~\ref{sec:conc} concludes the paper.

\section{Related Work}
\label{sec:rela}

Speaker augmentation techniques have been widely adopted in speaker recognition tasks~\cite{yamamoto2019speaker,chen2022build,wang2023speakeraugment}. 
Among these, speed perturbation is perhaps the most popular. It modifies the pitch and formants of a speech utterance while keeping the linguistic content unchanged. 
This simple yet effective strategy has been shown to be highly effective in speaker recognition under complex conditions~\cite{zhou2024comprehensive}.

Our work is built upon the speed perturbation approach, though we made a slight change to meet our goal of tackling the key problem of target confusion in target speaker extraction. Specifically, we designed a time-domain resampling and rescaling pipeline that produces speech of pseudo-speakers that differs from the original speech only in speaker traits. This allows us to construct hard samples that can be used to enforce the model to identify genuine speaker traits.


In speaker extraction, Li \textit{et al.}~\cite{li2024effectiveness} investigated data augmentation on enrollment speech by adding noise and reverberation. Our approach is speaker augmentation, which adds pseudo-speakers. Moreover, we emphasize constructing and exploiting ``hard samples'' to boost speaker discrimination, rather than simply increasing overall data volume. 
Notably, our speaker augmentation approach and conventional data augmentation methods are complementary and can be combined to gain further improvement, though we focus on a deep understanding of speaker augmentation in this paper and so leave the data-and-speaker augmentation as future work.

\section{Our Method}
\label{sec:meth}

\subsection{Speaker Augmentation}

\subsubsection{Step 1: Resampling}

We first perform speed perturbation (SP) via time-domain resampling~\cite{zhou2024comprehensive}. 
Given a speech signal $x(t)$, we modify its time axis using a perturbation factor $\alpha$, resulting in the output signal $y(t)$:
\begin{equation}
\label{eq:sample}
    y(t) = x\left(\alpha t \right).
\end{equation}

\noindent This time-domain modification induces a corresponding transformation in the frequency domain:

\begin{equation}
\label{eq:rcr}
X(f) \rightarrow \frac{1}{\alpha}X(\frac{1}{\alpha}f),
\end{equation}
\noindent where $X(f)$ and $\frac{1}{\alpha}X(\frac{1}{\alpha}f)$ denote the Fourier transforms of $x(t)$ and $y(t)$, respectively.

It can be observed that resampling stretches or compresses the spectrogram along both the temporal and the frequency axis.
Specifically, the fundamental frequency (F0) and the spectral envelope are raised (downsampling) or lowered (upsampling), 
leading to changes in pitch and vocal tract characteristics (formants). Consequently, the generated speech retains the same content but exhibits modified speaker traits, simulating a new different speaker.

\subsubsection{Step 2: Rescaling}

To restore the original speech tempo, we apply a time-domain rescaling using the WSOLA algorithm~\cite{verhelst1993overlap}implemented via the `sox tempo()' function.
WSOLA chops the audio into overlapping segments, shifts them in the time domain, and cross-fades them at points where the waveforms are most similar. The entire effect of the WSOLA algorithm is to adjust the speech tempo while maintaining the pitch, thus preserving the speaker traits.

Combining resampling and rescaling, the final augmented speech maintains the content, tempo, and prosody of the original speech while possessing altered speaker traits.

\subsection{Discussion}
\label{sec:discuss}

We argue that this speaker augmentation approach provides two significant advantages for E2E-SE systems.

\subsubsection{Increased Speaker Diversity}

Existing SE benchmark datasets, such as WSJ0-2Mix~\cite{hershey2016deep} and LibriMix~\cite{cosentino2020librimix}, 
contain a relatively small number of speakers (typically several hundred), significantly fewer than speaker recognition datasets, which often involve tens of thousands of speakers. 
With the limited number of speakers in the training data, it is hard to establish the speaker embedding space, which means that speaker generalizability cannot be ensured. 
By producing pseudo speakers, the speaker embedding space can be better constructed, which improves the speaker encoder's ability to generalize across unseen speakers. 

\subsubsection{Hard Sample Generation}

Conventionally, SE models are trained by mixed speech constructed by mixing two randomly selected utterances from two speakers. This naive mixing approach cannot prevent the model from learning spurious cues (e.g., textual similarity) rather than genuine speaker traits. For instance, the model may find that a particular phone sequence can be used to match the enrollment and target speech of the same speaker, and so use that cue to identify the target speech. This risk is particularly high when the number of speakers in the training data is limited, as the speaker generalizability cannot be ensured in this case, so non-speaker cues are likely to be identified as spurious. Unfortunately, SE falls in that case.

Our resampling-and-rescaling approach produces speech that retains the same content, tempo, and prosody as the original utterance and speaker trait is the only changed factor. 
When mixing the augmented and the original utterances, we obtain `hard samples' for which the model can only use speaker traits to extract the target speech. This forces the speaker embedding module and the extractor to capture and utilize genuine speaker characteristics. 

In summary, we argue that the effect of speaker augmentation is twofold: more speakers to establish a better generalizable embedding space and increased speaker discrimination by learning from hard samples. We will validate these two augmentations through extensive ablation studies in Section~\ref{sec:exp}.

\section{Experiments}
\label{sec:exp}

\subsection{Experimental Setup}

\subsubsection{Data}
We conduct experiments on two benchmark datasets: WSJ0-2Mix~\cite{hershey2016deep} and Libri2Mix~\cite{cosentino2020librimix}. 
Both datasets are used in their 8kHz versions.

\emph{WSJ0-2Mix:} This dataset is derived from the Wall Street Journal (WSJ0) corpus~\cite{garofolo2007csr} and 
consists of 2-speaker mixtures. 
The training set comprises 40,000 clean utterances from 101 speakers, 
while the standard test set includes 3,000 mixtures from 30 speakers.

\emph{Libri2Mix:} This dataset is based on the LibriSpeech corpus~\cite{panayotov2015librispeech} and contains 2-speaker mixtures. 
We use the \emph{train-100} subset, which consists of 27,800 clean utterances from 251 speakers. 
Evaluation is performed on two standard test sets: 
\emph{Libri2Mix clean} (average SNR = 0 dB) and \emph{Libri2Mix noisy} (augmented by noise signals from
the WHAM! dataset\cite{wichern2019wham}, average SNR = -2.2 dB), 
each containing 3,000 mixtures from 40 speakers.

For speaker augmentation, the perturbation factor $\alpha$ was selected from a limited set $\{0.8, 0.9, 1.0, 1.1, 1.2\}$, following the setting in~\cite{zhou2024comprehensive}. This expands the number of speakers fivefold. Unless explicitly specified, this fivefold augmentation is the default setting in our experiments.

\subsubsection{Models}
We evaluate our method using two state-of-the-art E2E-SE architectures: DPRNN~\cite{2020Dual} and SpEx+~\cite{ge2020spex+}.

\emph{SpEx+:} SpEx+ is an E2E-SE model that involves a speech encoder that processes speech signals with three-time scales: $L_1=2.5$ ms, $L_2=10$ ms, and $L_3=20$ ms. 
The speaker extractor consists of 4 stacked temporal convolutional networks (TCNs) modules, each module containing 8 TCN blocks. 
The speaker encoder comprises three stacked ResNet blocks, producing 256-dimensional speaker embeddings.

\emph{DPRNN:} DPRNN is a dual-path RNN-based model initially designed for speech separation. 
We adopt the DPRNN model from~\cite{pariente2020asteroid} to design our E2E-SE system. which demonstrates strong performance compared to prior E2E-SE system.
The speech encoder and decoder follow the configuration presented in~\cite{2020Dual}, with the time scale set to 2.5 ms. 
The speaker extractor employs a Bi-LSTM network with 128 memory units per direction and a bottleneck size of 64. 
The speaker encoder consists of three ResNet blocks, generating 256-dimensional speaker embeddings.

The loss function consists of two parts: (1) a speech reconstruction loss based on the scale-invariant signal-to-distortion ratio (SI-SDR)~\cite{luo2018tasnet}, 
to measure the distortion between the extracted and the clean target speech; 
(2) a speaker classification loss based on cross-entropy to ensure speaker discrimination.

\subsubsection{Training setup}
We employ a dynamic mixing strategy~\cite{alex2023data} to generate diverse training samples. 
For each target utterance $x_t$, a new mixture $y_t$ is dynamically generated by randomly mixing it with an interfering utterance. 
In addition, a different utterance $e_t$ from the same target speaker is randomly selected as the enrollment speech. 
Thus, the training samples are in the form of triplets $\{x_t, e_t, y_t\}$.

For WSJ0-2Mix, the SNR of the mixed speech is uniformly sampled between -5 dB and 5 dB. 
For Libri2Mix, we follow~\cite{cosentino2020librimix}, where the SNR of the `Libri2Mix clean' set follows a Gaussian distribution with a mean of 0 dB and variance of 16.81 dB, while the SNR of the `Libri2Mix noisy' set follows a mean of -2.2 dB with a variance of 12.96 dB. 

All the models are trained for a maximum of 200 epochs with an initial learning rate of 0.001. 
The learning rate is reduced by a factor of 0.5 if the validation loss does not reduce for two consecutive epochs. 
The Adam optimizer is used for training. 
The source code is publicly available\footnote{https://github.com/youzhenghai/TSEspkaug}.

\subsubsection{Evaluation metrics}

Two evaluation metrics are used to assess model performance.

First, we employ the scale-invariant signal-to-distortion ratio improvement (SI-SDRi)~\cite{luo2018tasnet} 
to measure speech extraction quality. 
A positive SI-SDRi value ($\text{SI-SDRi} \geq 0$) indicates that the extracted speech is closer to the clean target speech, 
whereas a negative value suggests that the target speaker's voice is not effectively extracted.

To further evaluate the model's ability to handle target confusion, 
we adopt the \emph{Negative SI-SDRi Rate} (NSR)~\cite{zhang2020x}, defined as:

\begin{equation}
\text{NSR} = \frac{1}{N} \sum_{k=1}^{N} \mathbb{I}(\text{SI-SDRi}^k < 0),
\end{equation}
\noindent where $k$ indexes the test samples, $N$ the total number of test samples, and $\mathbb{I}(\cdot)$ is an indicator function. $\mathbb{I}(\text{SI-SDRi}^k < 0)$ outputs 1 when $\text{SI-SDRi}^k \leq 0$, indicating the occurrence of target confusion in the $k$-th sample. Otherwise, the indicator function outputs 0.


\subsection{Basic Results}

We first evaluate the effectiveness of our proposed speaker augmentation method across the two E2E-SE frameworks and the two benchmark datasets. The experimental results are presented in Table~\ref{tab:basic}. It can be observed that our method consistently improves performance under all test conditions across both evaluation metrics, demonstrating its effectiveness.

More interestingly, the effect of speaker augmentation is more pronounced in challenging test conditions. 
For example, considering the SpEx+ model: in the Libri2Mix clean setting, integrating speaker augmentation relatively improves SI-SDRi by 3.78\% and reduces NSR by 6.57\%, while in the Libri2Mix noisy setting, speaker augmentation enhances SI-SDRi by 5.84\% and decreases NSR by 21.44\%. These results indicate that the generated pseudo-speaker data effectively enhances the model's robustness in complex scenarios. This supports our hypothesis that more speakers in the training data help improve the generalizability of the speaker embedding module. 

\begin{table}[th]
  \caption{Performance comparison across different models and datasets with and without speaker augmentation.}
  \label{tab:basic}
  \centering
  \resizebox{\linewidth}{!}{ 
  \begin{tabular}{ l  l  l  c  c }
    \toprule
    \textbf{Model} & \textbf{Dataset} & \textbf{Setting} & \textbf{SI-SDRi (↑)} & \textbf{NSR (↓)} \\
    \midrule
    \multirow{6}{*}{SpEx+} 
      & \multirow{2}{*}{WSJ0-2Mix} 
      & Baseline        & 16.91  & 2.35\% \\
      & & ~~+ SpkAug    & 17.34  & 1.10\% \\
    \cmidrule(lr){2-5}
      & \multirow{2}{*}{\makecell[l]{Libri2Mix\\clean}} 
      & Baseline        & 13.23  & 4.26\% \\
      & & ~~~+ SpkAug   & 13.73  & 3.98\% \\
    \cmidrule(lr){2-5}
      & \multirow{2}{*}{\makecell[l]{Libri2Mix\\noisy}} 
      & Baseline        & 10.96  & 4.85\% \\
      & & ~~~+ SpkAug   & 11.60  & 3.81\% \\
    \midrule
    \multirow{6}{*}{DPRNN} 
      & \multirow{2}{*}{WSJ0-2Mix} 
      & Baseline        & 18.62  & 3.78\% \\
      & & ~~~+ SpkAug   & 20.03  & 1.42\% \\
    \cmidrule(lr){2-5}
      & \multirow{2}{*}{\makecell[l]{Libri2Mix\\clean}} 
      & Baseline        & 14.34  & 4.00\% \\
      & & ~~~+ SpkAug   & 14.72  & 3.67\% \\
    \cmidrule(lr){2-5}
      & \multirow{2}{*}{\makecell[l]{Libri2Mix\\noisy}} 
      & Baseline        & 11.45  & 4.90\% \\
      & & ~~~+ SpkAug   & 11.98  & 4.38\% \\
    \bottomrule
  \end{tabular}
    }
\end{table}

\subsection{Effect of the Number of Augmented Speakers}

Next, we investigate how the number of augmented speakers affects the system performance. The results are presented in Table~\ref{tab:number}. One can observe that more pseudo-speakers lead to better performance. 
In particular, by comparing the results in the 1st row with those in the 5th and 6th rows, we observe that when 125 real speakers are expanded to 250 speakers (half real, half pseudo), the performance is close to that of using 251 real speakers (10.96 vs. 10.82/10.85 in SI-SDRi, 4.85\% vs. 4.87\%/4.91\% in NSR). 
This further indicates that the pseudo-speakers generated by speaker augmentation can simulate real speakers very well and look sufficient to help establish the speaker embedding space.

\begin{table}[ht]
  \caption{Performance of SpEx+ on Libri2Mix noisy with different numbers of (real/pseudo) speakers.}
  \label{tab:number}
  \centering
  \resizebox{\linewidth}{!}{ 
    \begin{tabular}{lllcc}
        \toprule
        \textbf{Setting} & \textbf{Spks} & \textbf{$\alpha$} & \textbf{SI-SDRi (↑)} & \textbf{NSR (↓)} \\
        \midrule
        Baseline     & 251 & 1.0 & 10.96 & 4.85\% \\
        ~~+ SpkAug   & 251 $\times$ 3 & 0.9, 1.0, 1.1 & 11.42 & 4.12\% \\
        ~~+ SpkAug   & 251  $\times$ 5 & 0.8, 0.9, 1.0, 1.1, 1.2 & 11.60 & 3.81\% \\
        \midrule
        Baseline     & 125 & 1.0 & 10.24 & 6.81\% \\
        ~~+ SpkAug   & 125 $\times$ 2 & 1.0, 1.1 & 10.82 & 4.87\% \\
        ~~+ SpkAug   & 125 $\times$ 2 & 0.9, 1.0 & 10.85 & 4.91\% \\
        \bottomrule
    \end{tabular}
  }
\end{table}

\subsection{Effect of the Hard Mixture Samples}

We further investigate the effect of hard samples. 
As discussed in Section~\ref{sec:discuss}, we hypothesize that speaker augmentation generates augmented data that, when mixed with the original utterances, forms more challenging training samples. 

To validate this hypothesis, we remove hard samples of different types from the training dataset, making the training set easier:

\begin{itemize}
    \item \textbf{Remove Same Tempo (S.T.) samples}: Augmented speech will only undergo resampling without rescaling, causing a misalignment in tempo.
    \item \textbf{Remove Same Content (S.C.) samples}: Augmented speech will no longer be mixed with the original speech.
    \item \textbf{Remove Same Speaker (S.S.) samples}: Augmented speech will no longer be mixed with speech from the original speaker.
\end{itemize}

We evaluate the performance changes when different types of hard samples are removed. 
In our experiments, we fix the total number of training samples and control batch composition to exclude specific hard samples. 
For S.C., it excludes mixtures of the same content, which account for about 1\% and for S.S., it excludes mixtures of the same speaker, which account for about 0.08\%. 
The results are summarized in Table~\ref{tab:hard}. Note that we only tested the condition with $\alpha$ set to 0.9 and 1.0, 
resulting in a twofold speaker expansion.

\begin{table}[ht]
  \caption{Effect of hard samples with SpEx+ on Libri2Mix noisy.}
  \label{tab:hard}
  \centering
  \resizebox{0.85\linewidth}{!}{ 
    \begin{tabular}{lcccc}
        \toprule
        \textbf{Setting} & \textbf{Spks}  & \textbf{SI-SDRi (↑)} & \textbf{NSR (↓)} \\
        \midrule
        Baseline        & 125             & 10.24    & 6.81\% \\
        ~~+ SpkAug      & 125 $\times$ 2  & 10.85    & 4.91\% \\
        ~~ ~~- S.C.      & 125 $\times$ 2  & 10.82    & 5.18\% \\
        ~~ ~~- S.S.      & 125 $\times$ 2  & 10.65    & 5.15\% \\
        ~~ ~~- S.T.      & 125 $\times$ 2  & 10.81    & 5.32\% \\
        ~~ ~~ ~~- S.C.   & 125 $\times$ 2  & 10.77    & 5.72\% \\
        ~~ ~~ ~~ ~~- S.S. & 125 $\times$ 2  & 10.57    & 6.20\% \\
        \bottomrule
    \end{tabular}
  }
\end{table}

Several key observations can be made:
\begin{itemize}
    \item Removing any of the three types of hard samples leads to a decline in system performance despite their small proportion, 
confirming that each contributes positively to model training.
    \item Among the three, removing S.S. results in the most significant drop in SI-SDRi, while its impact on NSR is less pronounced. 
Conversely, removing S.T. leads to the highest increase in NSR but has a relatively smaller impact on SI-SDRi. 
This suggests that different types of hard samples challenge the model in different ways.
    \item As more hard samples are removed, system performance degrades substantially, supporting our hypothesis that 
learning from such challenging samples enforces the model to learn genuine and discriminative speaker representations.
    \item These findings highlight the importance of incorporating diverse hard samples in model training. 
Further exploration of the underlying effects of each type of hard samples will be considered in our future work.
\end{itemize}

\subsection{Integration with Metric Learning}

Finally, we evaluate the complementarity of our speaker augmentation method with metric learning~\cite{zhao2022target}.
Specifically, in addition to the speech reconstruction loss and speaker classification loss, we add a triplet loss proposed in~\cite{zhao2022target} that 
aims to minimize the distance between the enrollment embedding and the target speaker embedding, while maximizing the distance between the enrollment embedding and the interfering speaker embedding.

The results are presented in Table~\ref{tab:fusion}. It can be seen that combining speaker augmentation and the triplet loss leads to accumulated performance gains. 
This indicates that speaker augmentation and metric learning are complementary and can be utilized together. 

\begin{table}[th]
  \caption{Performance of SpEx+ by combining speaker augmentation with metric learning.}
  \label{tab:fusion}
  \centering
  \resizebox{\linewidth}{!}{ 
  \begin{tabular}{llcc}
    \toprule
    \textbf{Dataset} & \textbf{Setting} & \textbf{SI-SDRi (↑)} & \textbf{NSR (↓)} \\
    \midrule
    \multirow{2}{*}{\makecell[l]{Libri2Mix \ Clean}}  
      & SpkAug        & 13.73  & 3.98\% \\
      & ~~ + Triplet Loss        & 13.79  & 3.73\% \\
        \midrule
    \multirow{2}{*}{\makecell[l]{Libri2Mix \ Noisy}}  
      & SpkAug        & 11.60  & 3.81\% \\
      & ~~ + Triplet Loss        & 11.67  & 3.58\% \\
    \bottomrule
  \end{tabular}
    }
\end{table}

\section{Conclusion}
\label{sec:conc}

In this paper, we propose a simple yet effective speaker augmentation strategy for end-to-end speaker extraction (E2E-SE). 
By applying a resampling-and-rescaling pipeline, we can generate vast pseudo-speakers, enriching speaker diversity in the training data. Moreover, the generated data preserved the same text, tempo, and prosody of the original utterances, 
resulting in hard samples that encourage the model to capture and utilize genuine speaker characteristics. We conducted extensive experiments across different model structures and datasets.
The experimental results demonstrated that the proposed method consistently enhances SI-SDRi while reducing target confusion. 
Furthermore, our approach is complementary to metric learning techniques, and their combination leads to additional performance improvement.

We admit the present work needs a more thorough investigation. In particular, the success is based on the present standard benchmark, where the number of speakers is limited, so speaker augmentation is expected to give a substantial contribution. To verify the true value of hard speaker augmentation in learning genuine speaker patterns, we need to test the proposal with larger datasets, e.g., those involving thousands of speakers. This needs to establish a new benchmark for the SE task.

\bibliographystyle{IEEEtran}
\bibliography{mybib}
\end{document}